\documentclass[twocolumn,prb, showpacs,preprintnumbers,amsmath,amssymb,superscriptaddress]{revtex4}

\usepackage{graphicx}
\usepackage{dcolumn}
\usepackage{longtable}%
\usepackage{bm}
\usepackage{color}

\begin{document}

\preprint{APS/123-QED}

\title{Disorder-driven electronic localization and phase separation in superconducting Fe$_{1+y}$Te$_{0.5}$Se$_{0.5}$ single crystals \\}

\author{S. R\"o{\ss}ler}
\email{roessler@cpfs.mpg.de}
\affiliation{Max Planck Institute for Chemical Physics of Solids,
N\"othnitzer Stra\ss e 40, 01187 Dresden, Germany}
\author{Dona Cherian}
\affiliation{Department of Physics, C.V. Raman Avenue, Indian
Institute of Science, Bangalore-560012, India}
\author{S. Harikrishnan}
\affiliation{Department of Physics, C.V. Raman Avenue, Indian
Institute of Science, Bangalore-560012, India}
\author{H. L. Bhat}
\affiliation{Department of Physics, C.V. Raman Avenue, Indian
Institute of Science, Bangalore-560012, India}
\affiliation{Centre for Liquid Crystal Research, Jalahalli, Bangalore-560013, India}
\author{Suja~Elizabeth}
\affiliation{Department of Physics, C.V. Raman Avenue, Indian
Institute of Science, Bangalore-560012, India}
\author{J. A. Mydosh}
\affiliation{Kamerlingh Onnes Laboratory, Leiden University, PO Box 9504, 2300 RA Leiden, The Netherlands}
\author{L. H. Tjeng}
\affiliation{Max Planck Institute for Chemical Physics of Solids,
N\"othnitzer Stra\ss e 40, 01187 Dresden, Germany}
\author{F. Steglich}
\affiliation{Max Planck Institute for Chemical Physics of Solids,
N\"othnitzer Stra\ss e 40, 01187 Dresden, Germany}
\author{S. Wirth}
\affiliation{Max Planck Institute for Chemical Physics of Solids,
N\"othnitzer Stra\ss e 40, 01187 Dresden, Germany}
\date{\today}




\begin{abstract}
We have investigated the influence of Fe-excess on the electrical transport and magnetism of  Fe$_{1+y}$Te$_{0.5}$Se$_{0.5}$ (y=0.04 and 0.09) single crystals. Both compositions exhibit resistively determined superconducting transitions ($T_{c}$) with an onset temperature of about 15~K.
From the width of the superconducting transition and the magnitude of the lower critical field $H_{c1}$, it is inferred that excess of Fe suppresses superconductivity. The linear and non-linear responses of the ac-susceptibility show that the superconducting state for these compositions is inhomogeneous. A possible origin of this phase separation is a magnetic coupling between Fe-excess occupying interstitial sites in the chalcogen planes and those in the Fe-square lattice. The temperature derivative of the resistivity $d\rho/dT$ in the temperature range $T_{c}$ $<$ $T$ $<$ $T_{a}$ with $T_{a}$ being the temperature of a magnetic anomaly, changes from positive to negative with increasing Fe. A  $\log 1/T$ divergence of the resistivity above $T_{c}$ in the sample with higher amount of Fe suggests a disorder driven electronic localization.
\end{abstract}

\maketitle

\section{Introduction} 
The discovery of superconductivity in the LnFeAsO (Ln = La, Ce, Pr, Sm) family of compounds with critical temperatures ($T_{c}$) up to 56~K \cite{kamiharajacs_130_3296_2008, chenprl_100_247002_2008, renepl_83_17002_2008} promoted an intense search for novel Fe-based superconductors with similar crystal structure.
Within a few months, several new superconducting phases were discovered. Among them, tetragonal FeSe has the nominally simplest crystal structure. 
It has no charge reservoir layer separating the Fe$_{2}$Se$_{2}$ layers and, hence, is considered as parent compound to all the Fe-based pnictide and chalcogenide superconductors.\cite{hsupnas_105_14262_2008}
The superconducting transition temperature ($T_{c}$) is found to be extremely sensitive to the Fe:Se ratio, and the highest $T_{c}\sim8.5$~K at ambient pressure is observed when the compound is closest to the stoichiometric composition.\cite{mcqueen2009extreme}
Nevertheless, application of pressure to FeSe raises $T_{c}$ as high as $\sim37$~K.\cite{mizuguchi2008superconductivity, medvedev2009electronic, margadonna2009pressure} 
By substituting Te for Se, $T_{c}$ is enhanced to $\sim15$~K for about 50\% Te doping.\cite{yehepl_84_37002_2008, fangprb_78_224503_2008}  
The end member 
Fe$_{1+y}$Te is non-superconducting and exhibits an incommensurate antiferromagnetic (AFM) order, coupled to a structural distortion near 67~K.\cite{li2009first}
The incommensurability $\delta$ in Fe$_{1+y}$Te can be easily tuned by the value of $y$, and the AFM order becomes commensurate for the samples close to the stoichiometric composition (i.e., $y \simeq$ 0).\cite{bao2009tunable}
In mixed Fe$_{1+y}$Te$_{1-x}$Se$_{x}$, the magnetic order is found to survive as short-range correlations for the samples with 0.25 $\leq x \leq$ 0.49 even in the superconducting state.\cite{bao2009tunable, wen2009short, Khasanov2009coexistance, lumsden2010evolution}  
More recently, pressure-induced static magnetic order is observed in superconducting FeSe.\cite{bendele2009pressure}
Density functional theory (DFT) calculations\cite{alaskaprb_78_134514_2008} on the stoichiometric end members FeSe and FeTe  indicate Fermi surface (FS) structures very similar to those in Fe-pnictides, where a spin-density-wave (SDW) ground state is obtained due to FS nesting.
%
In contrast to the DFT predictions, recent neutron diffraction studies demonstrate a composition-tunable ($\delta\pi$, $\delta\pi$) AFM order, which propagates along the diagonal direction of the Fe-square lattice in the $ab$-plane.\cite{bao2009tunable, li2009first} 
This is unlike Fe-pnictides, where the propagation vector of the SDW-type AFM order is along the ($\pi$, $0$) edge of the Fe-square lattice.\cite{de2008magnetic} 
In fact, a SDW gap was not observed in Fe$_{1+y}$Te, \cite{chen2009electronic, xia2009fermi} and FS nesting is not considered as the origin of magnetic order.
Alternatively, a fluctuating-local-moment scenario has been invoked in order to explain the unusual magnetic properties of Fe$_{1+y}$Te.\cite{ma2009first, turner2009kinetic, johannes2009microscopic}

At this point, it is worthwhile to mention that the phase diagram of the Fe chalcogenides is extremely complex.
In the case of FeSe, non-superconducting phases such as Fe$_{3}$Se$_{4}$, Fe$_{7}$Se$_{8}$, and hexagonal FeSe form in close proximity in the temperature-composition phase diagram.\cite{schuster1979transition} 
Hence, the tetragonal superconducting phase might contain these secondary phases in small quantities.
Further, the synthesis procedure is prone to oxygen contamination and thus producing unwanted phases such as Fe$_{2}$O$_{3}$ and Fe$_{3}$O$_{4}$.
All these phases are magnetic and detrimental to superconductivity.
Another crucial issue in the case of FeSe superconductors is the role 
played by excess of Fe.
It is exceedingly difficult to obtain perfectly stoichiometric Fe chalcogenides, and excess of Fe appears to be always present in synthesized compounds.\cite{hsupnas_105_14262_2008, fangprb_78_224503_2008, yehepl_84_37002_2008, mcqueen2009extreme, bao2009tunable} 
The excess Fe ions randomly occupy interstitial sites (designated as Fe(2) sites) in the chalcogenide layer.\cite{bao2009tunable, li2009first, liu2009charge} 
DFT calculations \cite{zhang2009density} focusing on Fe$_{1+y}$Te indicate that the excess of Fe occurs in the +1 valence state with each Fe$^{+}$ donating approximately one carrier to the FeTe layer. 
Further, Fe$^{+}$ is found to be strongly magnetic with a local moment of 2.4 $\mu_{B}$. 
These moments can be expected to couple with the magnetism of the FeTe sublattice resulting in a more complex magnetic order.  
It is predicted that, when FeTe is doped with Se, magnetism of interstitial Fe persists and results in a pair-breaking effect in the superconducting state.\cite{zhang2009density}
Indeed, recent experimental results clearly show suppression of superconductivity \cite{mcqueen2009extreme, janaki2009synthesis, liu2009charge} and localization effects \cite{liu2009charge, yadav2010effect} induced by excess Fe. 

Here we present resistivity, magnetization, linear and non-linear response of the ac-susceptibility of nominal Fe$_{1+y}$Te$_{0.5}$Se$_{0.5}$ single crystals for two different values of $y$. 
The results clearly demonstrate that Fe-excess causes a broadening of the superconducting transition, a phase separation in the superconducting state, and a localization of the charge-carrier in the normal state. 
\section{Experimental} 
\begin{figure}[tb]
\centering \includegraphics[width=7.5cm,clip]{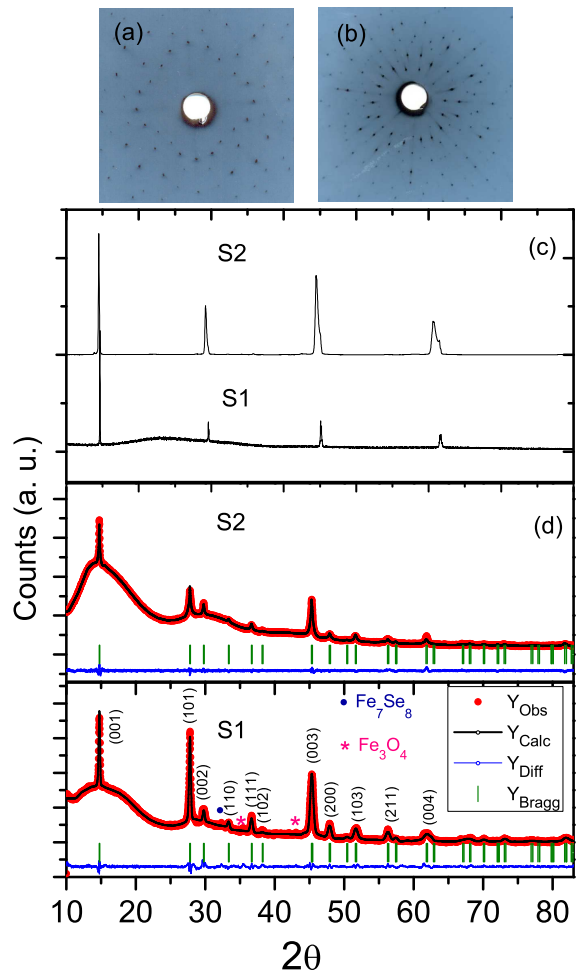}
\caption{(a)and (b) Laue diffraction patterns for S1 and S2, respectively.
(c) X-ray diffraction pattern of Fe$_{1.09}$Te$_{0.55}$Se$_{0.45}$ (S1) and Fe$_{1.04}$Te$_{0.52}$Se$_{0.48}$ (S2) single crystals displaying harmonic peaks corresponding to ($00l$) reflection (d)Powder XRD data of the crushed single crystals for samples S1 and S2.}
\label{fig.1}
\end{figure}
The single crystals used for the present investigation were grown using a horizontal Bridgman setup.
Appropriate quantities of iron (purity 99.9 \%), selenium (99.999 \%) and tellurium (99.999 \%) were mixed in a quartz ampoule in powdered form, evacuated to 10$^{-6}$ mbar, sealed and kept in a secondary quartz ampoule which is also evacuated and sealed. 
The ampoules were kept inside the Bridgman setup and the precursors were melted together at 950~$^{\circ}$C. 
Homogenization was done for 48 h by rotation of the melt in alternating clockwise and anticlockwise direction. 
After homogenization the furnace was translated at a rate of 9.2 mm/h so that a temperature gradient of 60~$^{\circ}$C/cm swept through the ampoule.
Finally, the ampoule was cooled to room temperature at a rate of 25~$^{\circ}$C/h.
Platelet-like single crystals of typical size of 5 mm $\times$ 4 mm with a thickness of 0.5$-$1~mm were obtained. 
The as-grown crystals can easily be cleaved along the $ab$-plane. 
Composition and elemental mapping along a certain direction was conducted by energy dispersive x-ray analysis (EDX).
The EDX compositions of the single crystals corresponding to different starting composition are listed in Table \ref{tab1}. 
\begin{table}[h!]
\caption{Nominal chemical composition (C$_{Nom}$), composition estimated from EDX (C$_{EDX}$), $c$-axis lattice constant ($c$-const), and label used for the two composition of Fe$_{1+y}$Te$_{0.5}$Se$_{0.5}$ single crystals used in this study. The $c$-axis lattice constants are estimated from the single crystal x-ray diffraction shown in Fig. \ref{fig.1}.}
\begin{tabular}{c|c|c|r}
\hline
\hline
C$_{Nom}$ & C$_{EDX}$ & $c$-const & label \\
 & &\AA& \\
\hline
Fe$_{1.25}$Te$_{0.5}$Se$_{0.5}$&Fe$_{1.09}$Te$_{0.55}$Se$_{0.45}$&6.032(8) &S1\\
Fe$_{1.05}$Te$_{0.5}$Se$_{0.5}$&Fe$_{1.04}$Te$_{0.52}$Se$_{0.48}$&6.109(3)&S2\\
\hline
\hline
\end{tabular}
\label{tab1}
\end{table}
%
%
%
%
%

%
The Laue photographs in Figs. \ref{fig.1}(a) and (b) indicate a good quality of the single crystals. 
The single-crystal  x-ray diffraction (XRD) data taken using Cu K$_{\alpha}$ radiation show, Fig. \ref{fig.1}(c), the harmonic peaks corresponding to the (00$l$) reflection and are comparable with those published by Yadav and Paulose. \cite{yadav2009upper}
%
%
In addition, we have conducted powder XRD on our samples, the results of which are presented in Fig \ref{fig.1}(d).
As is obvious from the comparison of Figs. \ref{fig.1}(c) and \ref{fig.1}(d) the single crystals can be much better characterized by powder XRD. This, however, requires crushing the single crystals and can, therefore, only be conducted once all other measurements are completed.
%
%
%
As identified in the Fig. \ref{fig.1}(d), sample S1 contains tiny peaks corresponding to small amounts of ($\leq 1\%$, see below) Fe$_{3}$O$_{4}$ and Fe$_{7}$Se$_{8}$ phases.
But these peaks are not detected in the XRD pattern of sample S2.
(Sample S2 might also contain these secondary phases below the detection limit of our powder XRD).
The structure refinement was performed by Rietveld method using the FULLPROF code.\cite{rod1993}
The samples have a tetragonal structure and belong to the $P4/nmm$ space group.
The lattice constants obtained from the refinement are $a$ = 3.7982(1), $c$ = 5.9990(4)~\AA~for sample S1 and $a$ = 3.7975(2), $c$ = 6.0031(5)~\AA~for sample S2.
These parameters are close to those reported by Sales $et~al$. \cite{sales2009bulk} for single crystals of similar composition. 
Transport and ac-susceptibility measurements were performed with a Quantum Design Physical Property Measurement System. 
Magnetization measurements were carried out by means of a SQUID magnetometer (Quantum Design). 
The measurements were conducted with current and field applied within the ab-plane.
\section{results and discussion} 
\begin{figure}[b!]
\centering \includegraphics[width=7.0cm,clip]{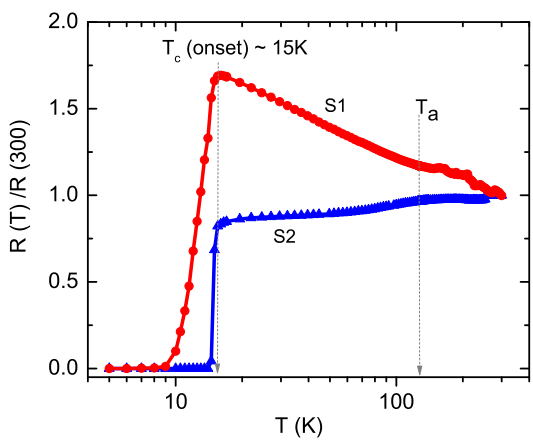}
\caption{Normalized in-plane resistance R(T)/R(300) as a function of temperature for Fe$_{1+y}$Te$_{1-x}$Se$_{x}$ single crystals on a semilogarithmic plot. For exact compositions, see Table \ref{tab1}. Note, for sample S1  R(T)/R(300) displays $-$log $T$ divergence below $T_{a}$ $\sim$ 130 K.This figure is highly similar to Fig. 1b in Ref. 25.}
\label{fig.2}
\end{figure}
\begin{figure}[tb!]
\centering \includegraphics[width=8.5 cm,clip]{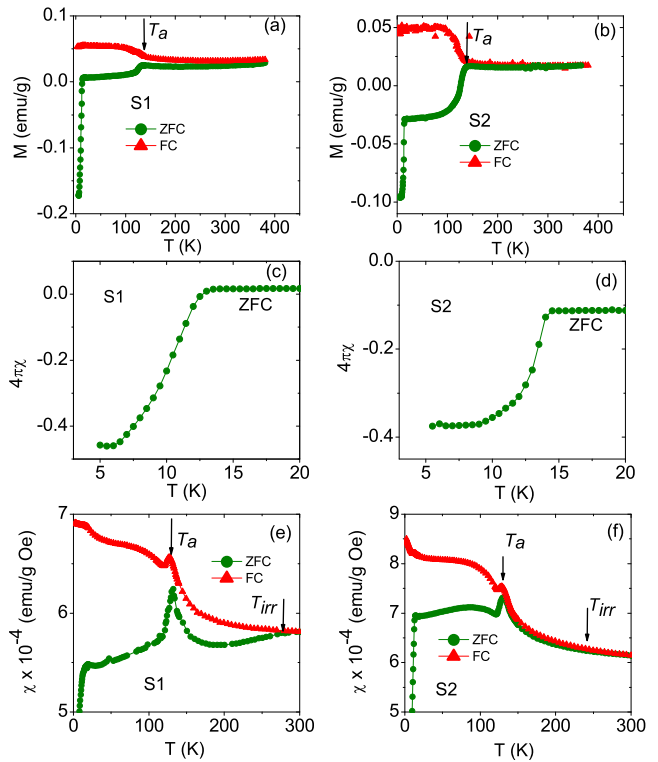}
\caption{(a) and (b) Zero-field-cooled (ZFC) and field-cooled (FC) magnetization as a function of temperature measured in a field of 30~Oe, applied parallel to the $ab$-plane, showing an anomaly at $T_{a} \sim$ 130 K and $T_{c} \sim$ 15~K. (c) and (d) ZFC  dc-susceptibility for $T$ $<$ 20 K. (e) and (f) ZFC and FC dc-susceptibility measured in a field of 1000 Oe, also displaying similar anomalies at $T_{a}$. An irreversibility observed between the ZFC and FC susceptibilities is marked by $T_{\mathrm{irr}}$.}
\label{fig.3}
\end{figure}

The influence of Fe excess on the electrical transport is immediately obvious in  Fig.~\ref{fig.2}, where the normalized resistance as a function of temperature for the two samples is plotted. 
The room temperature resistivity of samples S1 and S2 is about 0.9 and 0.6 m$\Omega$cm, respectively.
Both the samples show an onset of the superconducting transitions at around $T_{c} \sim$ 15~K, marked by the dotted vertical line in Fig.~\ref{fig.2}. 
However, the width of the superconducting transition increases from 1~K to 6~K as $y$ increases from 0.04 to 0.09.
Further, in the normal state, sample S2 displays a metallic behavior ($d\rho/dT$~\textgreater~0), whereas  a $\rho$ $\propto$ $\log 1/T$ divergence was observed for S1 below a temperature $T_{a} \sim$ 130~K.  
A similar divergence is also reported by Liu $et~al.$ for Fe$_{1.11}$Te$_{0.64}$Se$_{0.36}$ below 50~K. \cite{liu2009charge}
They also found a kink in resistivity at 120 K. 
The authors associated this kink with the magnetic anomaly observed earlier in polycrystalline 
samples. \cite{fangprb_78_224503_2008}
On the other hand, Janaki $et~al.$ \cite{janaki2009synthesis} attributed a similar anomaly observed around 125~K in the magnetization measurement of their polycrystalline samples to the Verwey transition of a Fe$_{3}$O$_{4}$ spurious phase within the grain boundaries.
%
In the present case, however, a $-$log $T$ divergence in $\rho(T)$ appears below  $T_{a}$, where an anomaly in the magnetization is observed (see Fig. \ref{fig.3}).
This suggests that the electrical transport is extremely sensitive to the disorder caused by unwanted secondary phases.
We note that a similar $-$log $T$ divergence was observed in the case of cuprates \cite{PhysRevLett.75.4662, PhysRevLett.77.5417, PhysRevLett.85.638} and 1111 Fe arsenides. \cite{riggs2009magnetic} This is ascribed to the onset of insulating behavior via disorder driven electron localization when superconductivity is suppressed by an external magnetic field.
%
%
%
%
%
%
%
%
%
%

Now we turn to the results of dc-magnetization and the ac-susceptibility, performed with the goal of establishing some evidence for the existence of local moments.
Figures \ref{fig.3} (a) and (b) show the zero-field-cooled (ZFC) and field-cooled (FC) magnetization for the samples S1 and S2 measured in a magnetic field of 30~Oe and in the temperature range 2~$-$~380~K.  
Although the ZFC magnetization is negative below the superconducting transition, positive values of FC magnetization are consistent with magnetic impurities.
Figures \ref{fig.3} (c) and (d) present the ZFC dc-susceptibility curve below 20 K. 
Clearly, the superconducting transition for S2 is sharper in comparison to that of S1.
%
%
\begin{figure}[tb!]
\centering \includegraphics[width=8.0cm,clip]{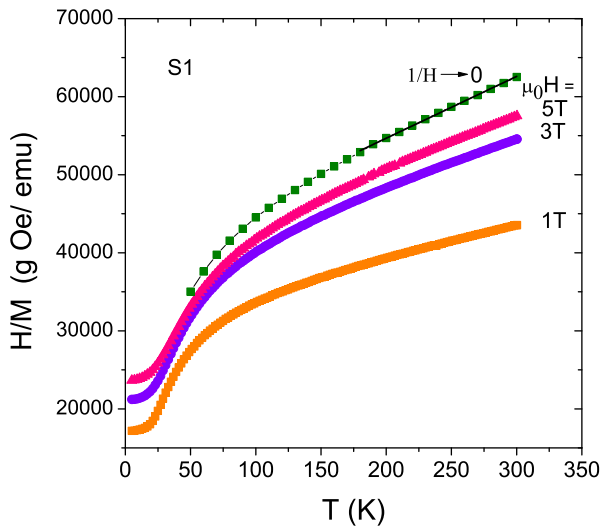}
\caption{Inverse magnetic susceptibility H/M(T) of sample S1 for different external magnetic fields and extrapolated values 1/H $\rightarrow$ 0, according to the Honda-Owen method. Line represents the Curie-Weiss fit.}
\label{fig.4}
\end{figure}
However, the fraction of the volume that is screened by superconducting currents estimated from the dimensionless dc-susceptibility is slightly less for sample S2, see Figures \ref{fig.3} (c) and (d). The full screening value is 4$\pi\chi = -1$.
The dc-susceptibilities measured in both FC and ZFC protocols with a field of 1 kOe are shown in Figs. \ref{fig.3} (e) and (f). Here, an irreversibility is clearly observed below $T_{\mathrm{irr}}$ of about 280~K for S1, and 260~K for S2 in the ZFC and FC susceptibility.
In addition to the superconducting transition, we observe an anomaly around $T_{a} \sim$ 130~K in both  samples.  
Comparing with Fig.~\ref{fig.2}, it can be noted that in the temperature dependence of the resistance, the poorer sample S1 obeys the characteristic $-$log $T$ divergence only below $T_{a}$, whereas the better sample S2 displays a broad maximum around $T_{a}$. 
There is no significant influence of the amount of Fe on the value of $T_{a}$. 
The change in the magnetization $\bigtriangleup M$ measured in a field of 1~kOe at $T_{a}$ for sample S1 is 1.6 $\times$ 10$^{-3} \mu_{B}$/f.u. and that for sample S2 is 1.0 $\times$ 10$^{-3} \mu_{B}$/f.u.
$\bigtriangleup M$ for Fe$_{3}$O$_{4}$ at the Verwey transition amounts to 0.25 $\mu_{B}$/f.u.\cite{paul2005magnetic}
If we attribute $\bigtriangleup M$ at $T_{a}$ in our measurements entirely due to the Verwey
transition of the secondary phase, then the estimated amount of Fe$_{3}$O$_{4}$ in sample S1 is $\sim$ 0.6~\% and that in sample S2 is $\sim$ 0.4~\%.
Note that similar anomalies have earlier been observed in polycrystalline Fe(Se$_{1-x}$Te$_{x}$)$_{0.82}$ where the value of $T_{a}$ varied with the amount of doping $x$.\cite{fangprb_78_224503_2008}
Neutron scattering studies on these samples did not reveal any magnetic or structural transition at this temperature.\cite{bao2009tunable}
However, a pronounced short-range quasielastic magnetic scattering at an incommensurate wave vector with a correlation length of 4~\AA~has been observed in a Fe$_{1.08}$Te$_{0.67}$Se$_{0.33}$ sample with optimal composition and highest $T_{c}$ $\sim$ 15~K.
The short-range quasielastic magnetic scattering was observed in both the normal and the superconducting states at the incommensurate wave vector (0.438, 0, $\frac{1}{2}$).\cite{bao2009tunable}
\begin{figure}[tb]
\centering \includegraphics[width=8.5cm,clip]{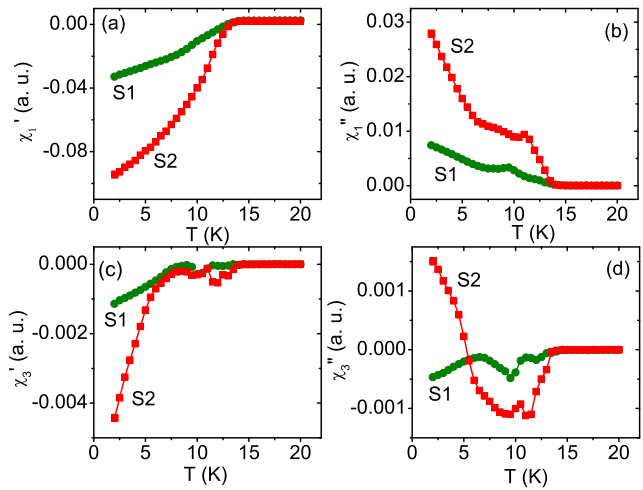}
\caption{AC-susceptibility as a function temperature measured in an ac field of 10 Oe and at a frequency of 1333 Hz for S1 and S2 samples. (a) real part $\chi_{1}'(T)$ of the linear susceptibility (b) imaginary part $\chi_{1}''(T)$ of the linear susceptibility (c)real part $\chi_{3}'(T)$ of the non-linear susceptibility, and (d) imaginary part  $\chi_{3}''(T)$ of the non-linear susceptibility.}
\label{fig.5} 
\end{figure}
%
%
%
%
%
%
Alternatively, neutron diffraction studies on FeSe$_{0.5}$Te$_{0.5}$ reported by Horigane $et~al.$ \cite{horigane2009relationship} showed that the width of the (200) peak changes below 125~K, suggesting a possible structural transition.
%
%
%
In order to unambiguously decide whether the anomaly at $T_{a}$ is associated with the Verwey transition of the Fe$_{3}$O$_{4}$ or whether it is an intrinsic property of the tetragonal Fe(SeTe), experiments which probe the sample properties on a more local scale are required.
%
%
In an attempt to extract the effective moments, the dc-susceptibility $\chi$ in the FC protocol is fitted to $\chi = \chi_{0}+C/(T-~\theta$) in the temperature range 180$-$300 K. 
Here, $\chi_{0}$ is the temperature-independent susceptibility arising from diamagnetic core, paramagnetic van Vleck contributions, diamagnetic Landau orbital and paramagnetic Pauli spin susceptibilities from conduction electrons. \cite{2009arXiv0911.4758Y, 2010arXiv1005.4392J} $C$ stands for the Curie constant and $\theta$ is the Weiss temperature. 
It is known that in Fe-containing samples, data analysis is often hampered by the contribution of a ferromagnetic impurity,\cite{PhysRevB.70.214418, 2009arXiv0911.4758Y} and the inverse susceptibility in the paramagnetic regime can thus be field dependent, see Fig. \ref{fig.4}.
Therefore, we utilized the Honda-Owen method \cite{honda1910} to eliminate the impurity contribution with the assumption that the magnetization of the ferromagnetic impurity saturates below 1~T.
In this method, the magnetic susceptibility M/H is plotted against 1/H for each temperature. 
A Curie-Weiss law can be fitted to the extrapolated values of the magnetic susceptibilities in the limit 1/H $\rightarrow$ 0 (Fig. \ref{fig.4}).
%
%
From the fit, we obtain $\chi_{0}=0.0019$ emu/g Oe, an effective moment of  $\mu_{\mathrm{eff}}$=1.49 $\mu_{B}$ and  $\theta=-50$~K for sample S1.
A similar approach for sample S2 provided $\chi_{0}=0.0017$ emu/g Oe, $\mu_{\mathrm{eff}}$=1.49 $\mu_{B}$, but $\theta=-88$~K.
%
%
%
%
%
A Curie-like behavior in Fe$_{1+y}$Te$_{1-x}$Se$_{x}$ has been reported by other research groups \cite{viennois2010effect, yadav2010effect, 2009arXiv0911.4758Y} as well and is attributed to Fe excess with localized moments.

%
%
%
%
%
%
%
%
%
%
%
\begin{figure}[tb!]
\centering \includegraphics[width=7.5 cm,clip]{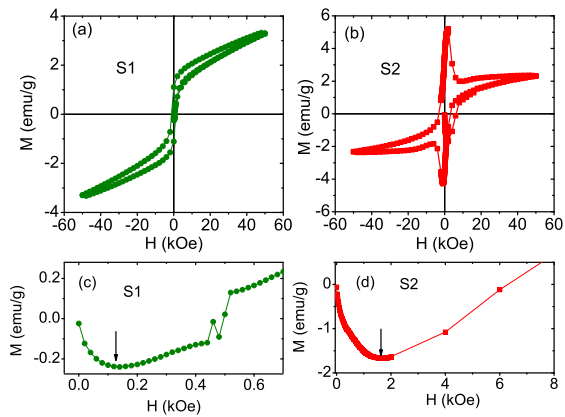}
\caption{Magnetization as a function of applied magnetic field at 2 K for samples (a) S1 and (b) S2. Panels (c) and (d) display the low-field initial magnetization curves for samples S1 and S2, respectively. The minima are marked by arrows (see text).}
\label{fig.6}
\end{figure}
\begin{figure}[t!]
\centering \includegraphics[width=6.0cm,clip]{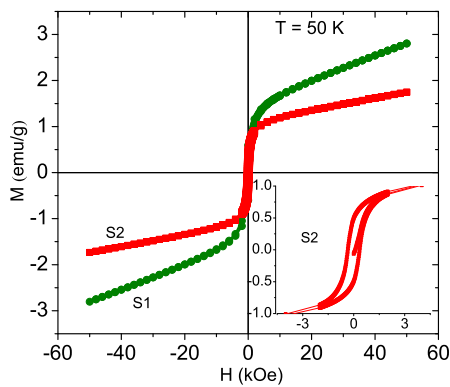}
\caption{Magnetization as a function of applied magnetic field at 50 K for samples S1 and S2. Inset enlarges the low field data for sample S2 showing a significant hysteresis.}
\label{fig.7}
\end{figure}
In order to further probe the superconducting state, we performed linear and non-linear ac-susceptibility measurements.
As this method gives more extensive information in the zero-field limit compared to the dc-magnetization, and  because frequency can be used as an additional tuning parameter, the method can provide insight into the nature of the transition not available, with the afore-mentioned techniques.\cite{PhysRevLett.10.212}
Further, the measurement of higher-harmonic susceptibility is even more useful because it only probes the non-linear magnetization.
The fundamental (linear) and higher-harmonic (non-linear) ac-susceptibility technique has extensively been used for characterizing the inhomogeneities in various superconductors including the high-$T_{c}$ cuprates.\cite{hein1992magnetic, PhysRevB.41.8937}
The technique is particularly useful in the case of Fe-based superconductors, where a phase separation of magnetic and superconducting entities is expected.\cite{bao2009tunable, wen2009short, Khasanov2009coexistance} 
In Figs. \ref{fig.5}(a)-(d), the real and imaginary parts of both the fundamental ($\chi_{1}$) and third-harmonic ($\chi_{3}$) are presented for the samples S1 and S2.
When a homogeneous sample goes through the superconducting transition, the real part of the linear susceptibility $\chi_{1}'$ always changes monotonically to the full screening value of $\chi= -1/4\pi$.
On the other hand, the imaginary part $\chi_{1}''$ in a homogeneous superconductor either changes monotonically or displays a peak and goes from its normal state value to substantially zero in the superconducting state.
%
%
%
%
Also, the magnitude of the third-harmonics $|\chi_{3}|$ = ($\chi_{3}'^{2}$+$\chi_{3}''^{2}$)$^{1/2}$ is taken to be proportional to $\chi_{1}''(T)$ \cite{ishida1981superconducting} and forms a peak in the temperature region of the superconducting transition.
In our samples, $\chi_{1}'(T)$ does not show full diamagnetic screening, Fig. \ref{fig.5}(a) , and $\chi_{1}''(T)$ displays a shoulder in Fig. \ref{fig.5}(b) rather than a peak below $T_{c}$.
Instead of a single sharp peak, $\chi_{3}'(T)$ and $\chi_{3}''(T)$ have double structures as shown in Fig. \ref{fig.5}(c) and Fig. \ref{fig.5}(d), respectively.
These are clear indications of a phase separation in the superconducting state.\cite{claus1992phase, asaoka1994homogeneity}
The phase separation into magnetic and superconducting phases is further revealed in the field dependence of magnetization ($M-H$) loops measured at 2~K, see Fig. \ref{fig.6}(a) and (b).
%
%
It is interesting to note that sample S1 contains a larger ferromagnetic component than sample S2, probably due to larger amounts of excess Fe.
As a result, the minimum in the initial magnetization curve which is related to the lower critical field $H_{c1}$, increases from $\sim$ 0.175 kOe  for S1 (Fig. \ref{fig.6}(c)) to $\sim$ 1.75 kOe for S2 (Fig. \ref{fig.6}(d)). 
This clearly indicates that the Fe excess suppresses the superconductivity.
Consequently, the mixed (vortex) state appears at a lower magnetic field in sample S1 with larger Fe excess.
The $M-H$ loops at 50~K in the normal state, Fig. \ref{fig.7}, displays a knee at low fields. 
The corresponding net moment estimated from the extrapolation of $M$ from the high fields to $H\rightarrow$~0 are 1.38~emu/g (0.046 $\mu_{B}$/f.u.) and 1.12~  emu/g (0.032 $\mu_{B}$/f.u.), for samples S1 and S2, respectively.
Further, a small hysteresis is seen even in S2, with lesser amount of excess Fe as shown in
the inset of Fig. \ref{fig.7}.
This indicates a ferromagnetic coupling, possibly originating from Fe-excess.
In fact, ferromagnetic behavior was earlier reported in FeSe thin films, \cite{feng2006ferromagnetic, wu2007nature} before superconductivity was discovered 
in these systems.
\section{Conclusions}
We investigated the influence of Fe excess on the magnetism and superconductivity in Fe-chalcogenide superconductors.
A ``metal''-``insulator'' transition is observed when the amount of Fe-excess is increased from $y=0.04$ to 0.09.
%
%
The ``insulating'' state is characterized by a $\log 1/T$ divergence, which suggests a magnetic impurity and/or disorder-driven electronic localization by the presence of Fe-excess.
This result is in accord with a scenario suggested by Liu $et~al.$ \cite{liu2009charge}
%
%
%
Evidence for a phase separation is provided by the non-linear ac-susceptibility for the compositions studied.
Our results clearly demonstrate that the physical properties of tetragonal Fe-chalcogenide are extremely sensitive to disorder and impurities.
%
Also, more experimental and theoretical studies are necessary to understand the nature of the couplings between interstitial Fe and the Fe in the Fe-square lattice.
%
\acknowledgments
The authors thank  L. Craco, C. Geibel and T. V. Ramakrishnan, for stimulating discussions. U. Burkhardt, C. Koz, and C. Shivakumara are gratefully thanked for their help in sample characterization. This work is partially supported by the German Academic Exchange Service (DAAD ID 50726385) and the Department of Science and Technology (DST-India). 
%

%

\end{document}